\documentclass[reprint, amsmath, amssymb, aps, prb, nomerge]{revtex4-1}
\pdfoutput=1

\usepackage[colorlinks, pageanchor=false]{hyperref} 
\usepackage{graphicx}
\usepackage{dcolumn}
\usepackage{bm}
\usepackage{siunitx}  
\setcitestyle{numbers, citesep = {,}}
\usepackage{easy-todo}
\usepackage{placeins}

\usepackage{color}
\newcommand{\citesq}[1]{[\cite{#1}]}

\begin{document}

\preprint{APS/123-QED}

\title{Non-trivial spatial dependence of the spin torques in L1$_0$ FePt-based tunnelling junctions}

\author{Mario Galante}%
\author{Matthew O. A. Ellis}
\author{Stefano Sanvito}
\email{sanvitos@tcd.ie}
\affiliation{%
 School of Physics and CRANN Institute, Trinity College Dublin, College Green, Dublin 2, Ireland
}%

\date{\today}

\begin{abstract}
We present an {\it ab-initio} study of the spin-transfer torque in a Fe/MgO/FePt/Fe magnetic tunnel 
junctions. We consider a FePt film 
with a thickness up to six unit cells, either in direct contact with the MgO spacer or with an intercalated 
ultra-thin Fe seed layer. We find that in the FePt layer the torque is not attenuated as strongly as in the case of pure Fe. Moreover, in FePt the torque 
alternates sign at the Fe and Pt atomic planes throughout the stack for all FePt thicknesses 
considered. Finally, when Fe is intercalated between MgO and L$1_0$-FePt, the torque is sharply 
attenuated and it is transferred to FePt only for a Fe seed layer that is less than two-atomic-planes thick.
We attribute these features to the different spatial profiles of the exchange and correlation field and 
the induced non-equilibrium spin accumulation. The calculated tunnelling magneto-resistance of 
the Fe/MgO/FePt/Fe junctions studied is enhanced with respect to the one of Fe/MgO/Fe, while it is 
reduced with Fe intercalation. Our work shows that L$1_0$-FePt junctions can be promising candidates 
for current-operated magnetic devices and that the magnetic texture at the atomic scale has an important 
effect on the spin transfer torque. 
\end{abstract}

\maketitle


\section{Introduction}

Magnetic random access memories are believed to be among the most promising candidates to deliver the future of scalable, 
non-volatile, rapidly accessible data storage. At the heart of these devices are magnetic tunnel junctions (MTJs), which 
store data on the relative orientation of the magnetisation vectors of two magnetic layers separated by an insulating barrier. 
Reading and writing such junctions can be efficiently performed by applying an electric current through the device; exploiting 
the tunnelling magneto-resistance (TMR)~\citesq{TMR} effect for reading and using spin-transfer torque (STT)~\citesq{STT} 
to write. STT arises when a current passes across two ferromagnets having different magnetisation directions and it is caused 
by the transfer of angular momentum between the two mediated by the current. The conduction electrons become spin polarised 
by passing through the first magnetic layer and their angular momentum is then transferred to the second. The ideal insulating 
barrier acts as a spin-filter maximising the spin-polarisation of the current and hence the torque.

Optimising the device structure to achieve low write currents is an important challenge in realising the potential of these devices.
Whilst early demonstrations of MTJs focused on devices with in-plane layers magnetisation, the write current can be reduced 
significantly by adopting an out-of-plane geometry, where the magnetisation direction of both layers is oriented normally 
to the barrier interface. In junctions with this configuration, known as perpendicular MTJs (pMTJs), a large perpendicular magnetic 
anisotropy (PMA) is required to overcome the shape anisotropy of the thin film and enforce thermal 
stability in scalable devices.

State-of-the-art devices are based upon CoFeB/MgO thin films, which can reach a TMR of up to 604\% at room temperature 
and 1144\% at low temperature~\citesq{Ikeda2010}. Furthermore, a large PMA has been observed at the CoFeB/MgO interface which is sufficient to achieve a perpendicular geometry in ultra-thin layers~\citesq{Worledge2011a}. Alternatively, L$1_0$ FePt is 
a popular material choice for high-density magnetic recording, since it has a large magneto-crystalline PMA, $K_u = \SI{7e6}{Jm^{-3}}$, 
allowing stable grain sizes down to a few nanometres~\citesq{Weller2000}. Despite the large uniaxial anisotropy, switching has been 
observed in FePt/Au giant magneto-resistance pillars with the aid of an applied magnetic field~\citesq{Seki2006}. Theoretical 
calculations of a FePt/MgO MTJ predicts a TMR of 340\% for a Fe terminated interface~\citesq{Taniguchi2008}.

Unfortunately, growing FePt/MgO devices can be challenging since the lattice mismatch between L1$_0$ FePt and MgO 
is large, $\sim 8.5\%$~\citesq{Cuadrado2014a}. This may cause issues during the growth process, such as the inability of
preserving the epitaxy across uneven layers. Strain can also cause a significant change in the magnetic properties of the 
FePt layer. In particular calculations have shown that a strain of 4\% can reduce the PMA to about 10\% of its original 
value~\citesq{Seki2006}. Practically, such strain can be reduced by inserting a seed-layer with a more amenable lattice 
constant at the MgO/FePt interface.
 
In this work we investigate a series of FePt/MgO-based pMTJs in order to establish their potential for future device applications. 
We utilise \emph{ab-initio} models to calculate the spin-transfer torque and the TMR for a range of FePt-based MTJ structures. 
We begin by detailing our computational method, before presenting results on the atom-resolved STT in the zero-bias 
limit for an Fe/MgO/Fe junction. This has an electronic structure analogous to that of CoFeB-based MTJs and hence provides 
a useful starting point for the discussion. We then continue with the analysis of the torque acting on the MTJs with FePt/Fe 
free layers and with a thin Fe seed layer intercalated at the MgO interface. In this case we vary the thickness of both the 
FePt layer and the seed layer (including the case where there is no seed layer). We find that a MgO/FePt interface yields a 
STT that decays more slowly in the free layer than in the MgO/Fe case, while the insertion of a Fe seed layer produces results similar 
to the FePt-free case. We then present the outcome of our TMR calculations and the STT acting on the Fe reference layer for 
some representative cases. Finally we replace the Fe atoms in the seed layer with Ni. This provides a comparison and 
helps us to formulate an argument about the origin of the spatial dependence of the STT.

\section{Computational Method}

Our approach for calculating the spin-transfer torque follows the prescription provided by Haney et al. in reference~\citesq{Haney2007} 
and is based on isolating the transport (non-equilibrium) contribution to the density matrix from the equilibrium part. The 
influence that an electric current has on the system can be estimated from first principles by combining density functional theory 
and the non-equilibrium Green's functions method for transport (DFT+NEGF). All calculations have been performed with the 
{\sc Smeagol} code~\citesq{Rocha2005,Rocha2006,Rungger2008a,Rungger2009}, which implements the DFT+NEGF scheme
within the numerical atomic orbital framework of the {\sc Siesta} package~\citesq{Soler2002}. 
\begin{figure}[t!]
 \begin{center}
\includegraphics[width=\columnwidth]{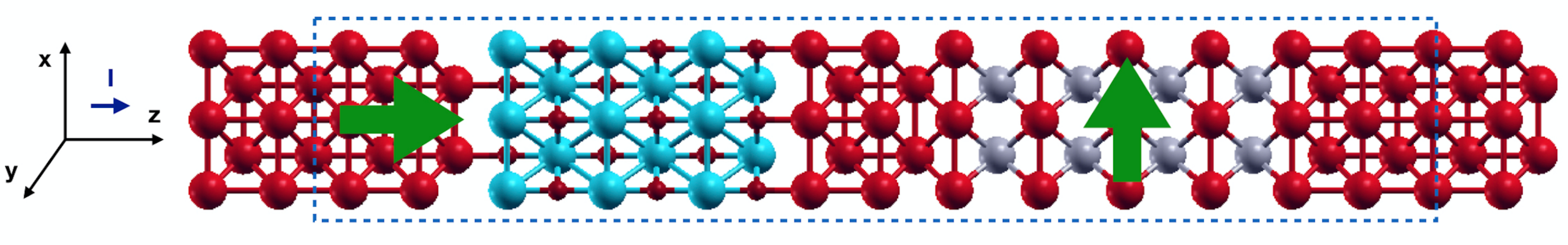}
  \caption{(Colour Online) Set up for a quantum transport calculation of a Fe/MgO/Fe/FePt/Fe junction. The dashed rectangle 
  delimits the scattering region from the leads. The green arrows indicate the different direction of the magnetisations of the 
  magnetic layers at the left- and right-hand side of the insulating barrier. The coloured spheres represent atoms of different 
  species: Fe atoms are in red, Pt in grey, O in light blue, Mg are small red spheres.}\label{setup}
 \end{center}
\end{figure}

The system set up for the quantum transport calculation sandwiches the magnetic tunnel junction between two semi-infinite leads 
(see Fig.~\ref{setup}). These are assumed to be made of bulk material and at equilibrium. Note that a certain portion of the 
electrodes has to be included in the scattering region in order to ensure the continuity of the electrostatic potential. Here the 
magnetisation of the reference or fixed layer, ${\bf M}_\text{ref}$, is considered to be magnetised along $z$ (the transport direction) 
and the one of the free layer, ${\bf M}_\text{free}$, along $x$, so that the two form a $\pi/2$ angle. A voltage is applied in such a 
way that the electron flux is flowing along the stacking direction, $z$, in our convention from the reference layer to the free one. 

The component of the torque vector, ${\bf T}$, which is responsible for the switching between the parallel and the anti-parallel 
magnetisation configurations is the one that lies in the plane defined by ${\bf M}_\text{free}$ and ${\bf M}_\text{ref}$, 
namely the $x-z$ plane. In the free layer this component coincides with $T_z$, which is the main focus of our study.
In order to reduce the computational costs, we limit our analysis to the torque response to a small bias, the \emph{torkance}, 
meaning that all calculations are performed in the linear response approximation. At an atom $a$ in the free layer the torkance 
is defined as
\begin{equation}\label{torque:def}
  \tau_z^a \equiv \frac{dT^a_z}{dV}\Big|_{V=0} = \frac{1}{2}\text{Re}\sum_{i\in a}\sum_j \left(\boldsymbol\Delta_{ij}\times\frac{d\boldsymbol m_{ji}}{dV}\Big|_{V=0}\right)_z\:,
\end{equation}
and this can be estimated with a zero-bias calculation. Here $\boldsymbol\Delta$ denotes the exchange and correlation 
field, namely the derivative of the exchange and correlation energy, $E_\mathrm{XC}$, with respect to the magnetisation 
density, $\boldsymbol m$, $\boldsymbol\Delta={\delta E_\mathrm{XC}}/{\delta {\boldsymbol m}}$. Thus, the derivative 
of $\boldsymbol m$ with respect to voltage embodies the spin contribution due to the rearrangement of the electronic 
population under non-equilibrium conditions. Henceforth this will be referred to as the non-equilibrium spin density or
the spin accumulation. As such, the torque is the result of the interaction between the internal static field $\boldsymbol\Delta$ 
and the non-equilibrium spin density generated by the current flow. Further details on the calculation of the spin-transfer 
torque and the torkance can be found in Refs.~\citesq{Stamenova2016, Ellis2017}.

A series of junctions are constructed, all having a barrier of 6 MgO layers sandwiched between two semi-infinite leads of bulk 
{\it bcc} Fe oriented along the (001) direction. Periodic boundary conditions are applied in the plane perpendicular to the transport, 
as a result of the perfect epitaxy of the junction. The in-plane lattice constant is taken to be $a_\text{Fe}=\SI{2.866}{\angstrom}$ 
throughout the system. The out-of-plane lattice constants of the remaining materials were chosen according to information provided 
in reference~\citesq{Kohn2013,Cuadrado2014a}, in particular $c_\text{MgO}=4.05/\sqrt{2}$ $\SI{}{\angstrom}$, 
$c_\text{FePt} = \SI{1.737}{\angstrom}$. The same studies assess that the most stable interfacial configuration is made of a 
Fe-terminated FePt surface on top of O (Fe) for the FePt/ MgO (FePt/Fe) interface, with an inter-plane distance of $\SI{2.2}{\angstrom}$ 
($\SI{1.585}{\angstrom}$). The accuracy of such estimates was found satisfactory by relaxation of the different structures. The local 
spin density approximation (LSDA) for the exchange correlation potential was adopted.  A real-space mesh cut-off of \SI{900}~{Ry} along 
with a 15$\times$15 $k$-point mesh in the plane orthogonal to transport were found to yield converged results. 
We adopted double $\zeta$ polarised orbitals for each atomic species and the convergence of the radial cut-offs was verified 
by comparing the band structure of bulk materials with the result of all-electron calculations. Since the introduction of spin-orbit 
coupling effects did not yield a sizeable change to our calculated torques we have omitted relativistic corrections.

\section{Results}
\begin{figure}[!t]
\includegraphics[width=.5\textwidth]{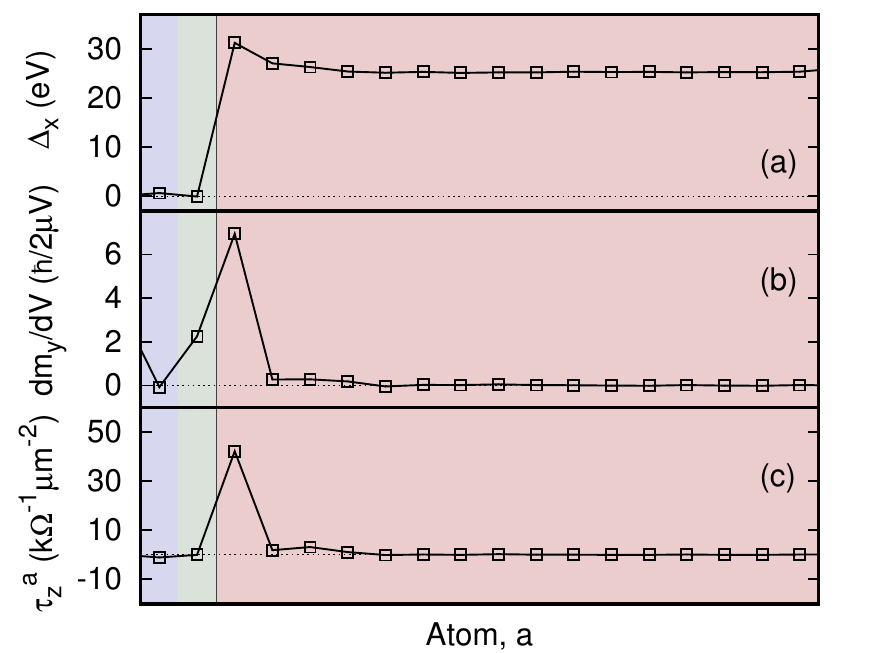}
\caption{(Color Online) Real space profiles of the relevant components of (a) the exchange and correlation field, 
$\boldsymbol\Delta$, (b) the non-equilibrium spin density, $\mathrm{d}\boldsymbol{m}/\mathrm{d}V$, and (c) the 
torkance, $\boldsymbol\tau$,  per unit $\mu_\mathrm{B}/e$ and area acting on the {\it bcc} Fe free layer. The coloured 
background indicates the atomic species in the stack: red for Fe, blue for O, green for Mg.}\label{FeMgO}
\end{figure}

We begin by examining the properties of a Fe/MgO/Fe MTJ to later discuss their modification upon the introduction 
of a FePt layer. As shown in Eq.~(\ref{torque:def}), the torkance is given 
by the vector product of the exchange and correlation field and the non-equilibrium spin density. Since the free layer is 
magnetised in the $x$-direction and within the LSDA the exchange and correlation field is proportional and locally parallel 
to the magnetisation, the only relevant components to the torkance are $\Delta_x$ and $\mathrm{d}m_y/\mathrm{d}V$. 
These two components and the resulting torkance, $\tau_z$, are shown in Fig.~\ref{FeMgO}. 

In general, $\Delta_x$ peaks at the Fe/MgO interface and then presents small oscillations with the period of the interlayer 
Fe separation, $a_\text{Fe}$. Such profile does correlate with the real space profile of the equilibrium magnetic moment 
(not displayed), which is also enhanced at the Fe/MgO interface. In contrast, the non-equilibrium spin density [panel (b)] has an 
appreciable magnitude only in the region around the Fe/MgO interface. This decays in the Fe layer and is almost fully attenuated 
a few monolayers from the interface. Such behaviour will later be compared with that in FePt and in Ni. Finally note that 
there is an appreciable non-equilibrium spin density also in the MgO, although it does not contribute to the torkance since the 
exchange and correlation field vanishes in absence of a local magnetization [see panel (a)].

If we now consider the torkance we note that this is sharply peaked at the Fe/MgO interface and is attenuated in the Fe layer 
at the same speed of the non-equilibrium spin density. In fact, for this Fe/MgO/Fe case the spatial dependance of the torkance
resembles closely that of the spin accumulation, given the fact that the exchange and correlation field has little spatial dependence
in Fe. Let us remark, however, that the point-by-point vector product of the quantities in panels (a) and (b) does not give the 
torkance in panel (c), since the sum of the products of the matrix elements does not equal the product of the sums [namely,
$\sum_{ij} \left(\boldsymbol\Delta_{ij}\times\frac{d\boldsymbol m_{ji}}{dV}\right)\ne
\left(\sum_{i}\boldsymbol\Delta_{ii}\right)\times\left(\sum_{i}\frac{d\boldsymbol m_{ii}}{dV}\right)$ - see formula (\ref{torque:def})].  

\begin{figure*}[!t]
\begin{center}
	\includegraphics[width=17.78cm]{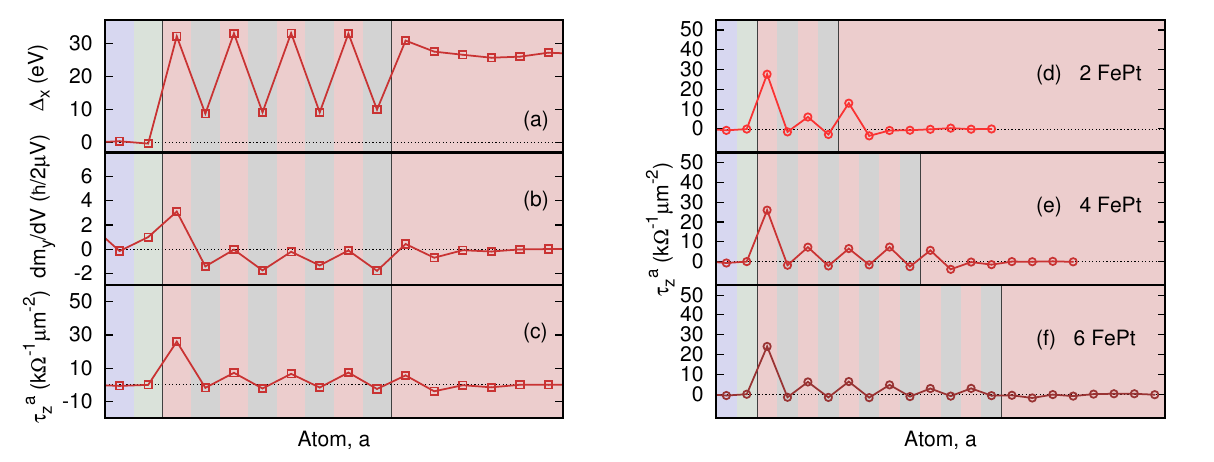}
\end{center}
\caption{(Color Online) Study of the torkance in a FePt/Fe free layer. Left panel: the relevant components of (a) the exchange 
and correlation field, $\boldsymbol\Delta$, (b) the non-equilibrium spin density, $\mathrm{d}\boldsymbol{m}/\mathrm{d}V$, 
and (c) the torkance per unit of $\mu_\mathrm{B}/e$ and area, $\boldsymbol\tau$. Right Panel: comparison of the torkance per 
unit $\mu_\mathrm{B}/e$ and area of MTJs with (d) 2, (e) 4 and (f) 6 FePt unit cells. In both subfigures the coloured background 
indicates the atomic species: red for Fe, grey for Pt, blue for O, green for Mg.}\label{FePt:noSL}
\end{figure*} 

We now explore the effects of inserting a layer of FePt at the MgO/free layer interface. Figures~\ref{FePt:noSL}(a)-(c) show, 
as with the case of the Fe/MgO/Fe MTJ, the relevant components of $\boldsymbol\Delta$ and $\mathrm{d}\boldsymbol{m}/\mathrm{d}V$
contributing to the total torkance along $z$ for a FePt layer 4-unit-cell thick. From panel (a) it is clear that the exchange
and correlation field is enhanced at the Fe sites, and also finite at the Pt ones. This is because in L$1_0$ FePt there 
is an induced magnetic moment on the Pt ions (this is about $0.4~\mu_\mathrm{B}$ as calculated from the M\"ulliken population 
analysis), in agreement with previous \emph{ab-inito} calculations \citesq{Cuadrado2014a}. The oscillations in the $\Delta_x$ profile 
remain constant in the FePt layer without any sign of decay, and then in the Fe layer the $\Delta_x$ profile returns to resemble 
the one observed before in Fig.~\ref{FeMgO}. Note that $\boldsymbol\Delta$ is an equilibrium property, which essentially 
depends on the presence of an exchange splitting in a given material. As such one does not expect a decay of $\boldsymbol\Delta$
unless there is a decay in the magnetisation.

 \begin{figure*}[!t]
 \begin{center}
	 \includegraphics[width=17.78cm]{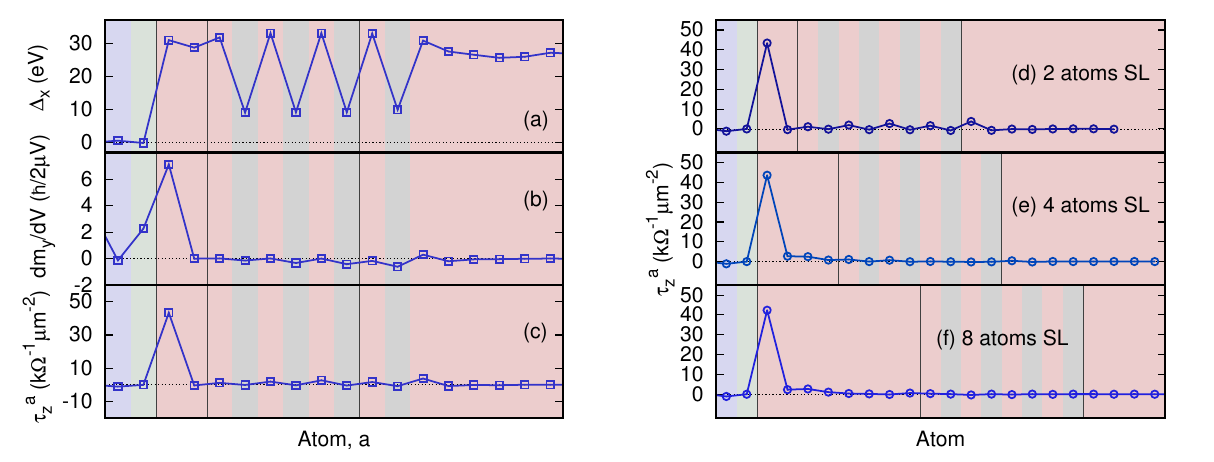}
\caption{(Color Online) Study of the torkance in a Fe/FePt/Fe free layer made of 4 FePt monolayers and a variable number of 
Fe monolayers inserted between MgO and FePt. Left panel: the relevant components of (a) the exchange and correlation field, 
$\boldsymbol\Delta$, (b) the non-equilibrium spin density, $\mathrm{d}\boldsymbol{m}/\mathrm{d}V$ and (c) the torkance per 
unit of $\mu_\mathrm{B}/e$ and area, $\boldsymbol\tau$. Right Panel: comparison of the torkance per unit $\mu_\mathrm{B}/e$ 
and area of MTJs with with a seed layer comprising (d) 2, (e) 4 and (f) 8 Fe monolayers. In all panels the coloured background 
indicates the atomic species: red for Fe, grey for Pt, blue for O, green for Mg.
}\label{FePt:FeSL}
 \end{center}
\end{figure*}

In contrast to the pure Fe case, the non-equilibrium spin density has lower intensity in FePt than in Fe but a significantly less 
attenuated decay [panel (b)]. The total non-equilibrium spin density shows regular oscillations within the FePt layer, whilst it is 
enhanced at both the FePt/Fe and the MgO/FePt interfaces, and then vanishes within a few unit cells of the Fe lead. 
Furthermore we observe that $\mathrm{d}{m_y}/\mathrm{d}V$ in Pt has opposite sign with respect to that of the first Fe layer 
in contact to MgO. Finally, the torkance [panel (c)] is again peaked at the interface with MgO but its strength is reduced in 
comparison to that computed for the Fe/MgO/Fe MTJ with the same MgO thickness. Within the FePt layer the torkance does 
not attenuate as in Fe but persists to reach the Fe-only side of the free layer. Most interestingly the torkance has an oscillatory 
behaviour in FePt, presenting small negative values at the Pt layers and positive at the Fe ones. Such oscillations are common 
in antiferromagnets~[\cite{Stamenova2016}] and here are observed also in a ferromagnet with non-trivial magnetic texture. It 
is also interesting to note that, despite the larger spin accumulation at Pt sites, the resulting torque is smaller than 
that at the Fe ones. This is due to the fact that the exchange and correlation field in Pt is significant weaker than in Fe (because
the magnetisation is smaller). 

The persistence of the torkance in the FePt layer remains as we change the FePt thickness, $n_\text{FePt}$ (number of unit cells). 
This can be seen in the panels (d)-(f) of Fig.~\ref{FePt:noSL}. For a thin layer [panel (d)] the torque is enhanced at the FePt/Fe 
interface, while it is attenuated for all the other cases [e.g. see $n_\text{FePt}=6$ in panel (f)]. Furthermore, for all the thicknesses 
considered the torkance remains strikingly positive at all the Fe atomic planes of FePt, while it is small and negative at the Pt ones. 
Moreover, the intensity of the peak at the MgO/FePt is not modified by the increase in thickness.

\begin{figure}
 \begin{center}
  \includegraphics[width=\columnwidth]{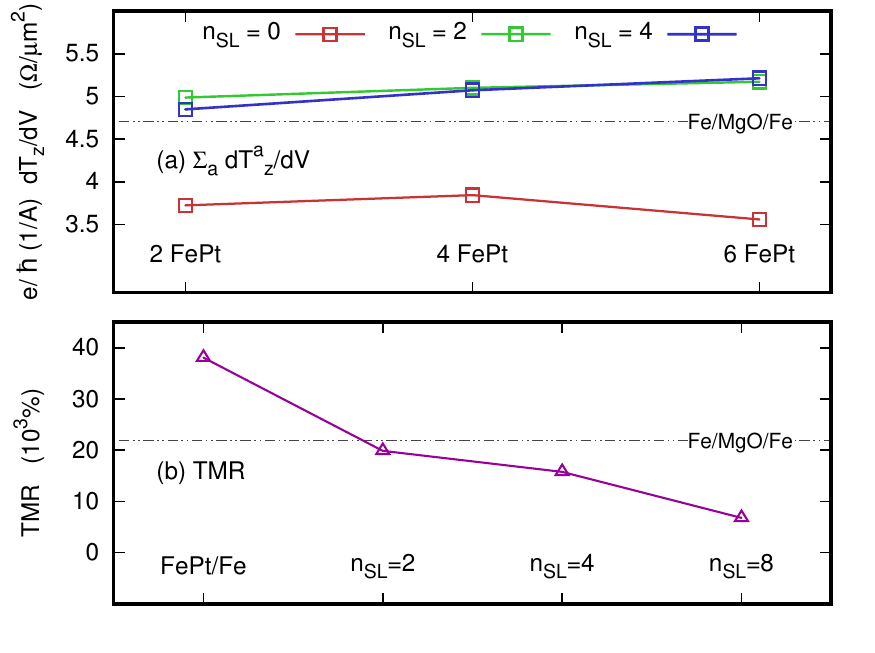}
\caption{(Colour Online) Panel (a): the total torkance per unit $\mu_\mathrm{B}/e$ and area acting on the free layer of 
Fe/MgO/Fe/FePt/Fe junctions with 2, 4 and 6 FePt monolayers and a Fe seed layer of $n_\mathrm{SL}=0, 2, 4$ atomic 
planes. Panel (b): the calculated TMR in Fe/MgO/FePt/Fe and Fe/MgO/Fe/FePt/Fe junctions with $n_\mathrm{SL}=2, 4, 8$ 
and 4 FePt unit cells. In both graphs the black dashed line represents the same quantity calculated for the Fe/MgO/Fe 
junction.}\label{TotT+TMR}
 \end{center}
\end{figure}

Although Fe/MgO/FePt/Fe junctions provide an interesting case of study, the significant lattice mismatch between MgO and 
L1$_0$ FePt ($\sim$ 8.5\%) makes their experimental realisation troublesome. 
This problem may be overcome by inserting a compatible seed layer at the MgO/FePt interface. Hence, we have 
analysed the influence of incorporating a thin Fe seed layer (SL) between the MgO and the FePt, keeping the thickness of the FePt 
layer constant at 4 unit cells. The Fe SL has different effects depending on its thickness (see Fig. \ref{FePt:FeSL}). We  
notice from panel (a) that the exchange and correlation field profile in FePt is analogous to the previous case (since the equilibrium 
magnetisation profile is also unchanged), while $\Delta_x$ is almost constant in the seed layer. The non-equilibrium spin density 
still oscillates in FePt, although the amplitude of such oscillations is much smaller than that obtained in absence of the SL. 
Consequently, the torkance [panel (c)] is peaked at the MgO/Fe interface with the SL and its intensity is comparable to that observed 
for the Fe/MgO/Fe case (see figure~\ref{FePt:noSL}). The torkance, however, is not exactly zero away from the SL, in particular on 
the Fe atoms of FePt and at the FePt/Fe interface. This does not happen for thicker Fe SLs [panels (e) and (f)], for which the total 
torkance decays before reaching the interface with FePt. In general, however, the main effect of the seed layer is to suppress
the persistence of the torkance in FePt, so that all the angular momentum transfer takes place in the seed layer.

\begin{figure}[!t]
 \begin{center}
  \includegraphics[width=\columnwidth]{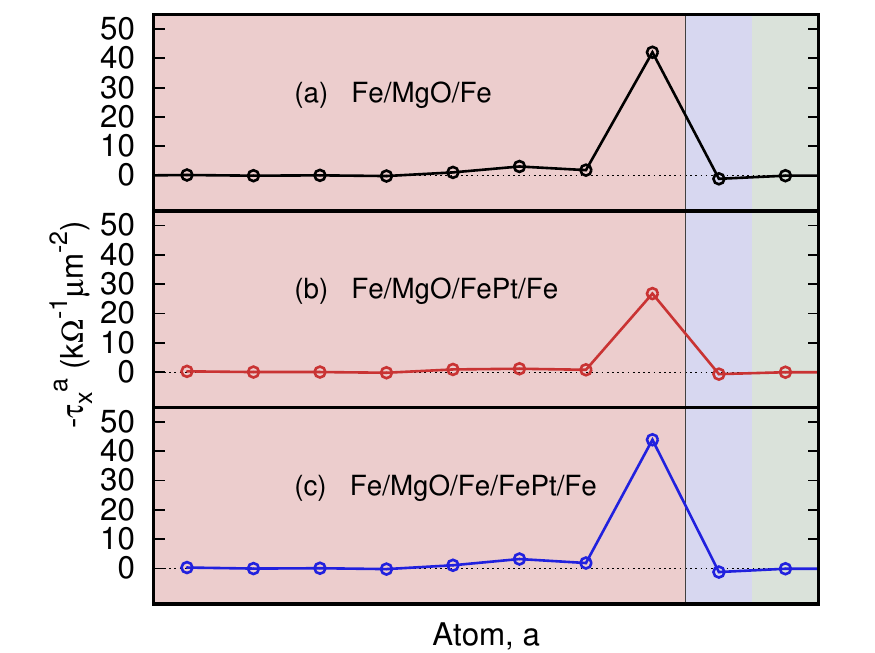}
\caption{(Colour Online) Torkance per unit $\mu_\mathrm{B}/e$ and area acting on the reference layer of Fe/MgO/Fe (a), 
Fe/MgO/FePt(4)/Fe (b) and Fe/MgO/Fe(2)/FePt(4)/Fe (c) MTJs. The coloured background indicates the atomic species to 
which each point corresponds to: red for Fe, blue for O, green for Mg. }\label{RefL}
 \end{center}
\end{figure}

We now move to analyse the total torkance and the TMR of each junction. Figure \ref{TotT+TMR}(a) shows the total torkance 
integrated over the free layer, $\tau_z^\mathrm{tot}=\sum_a^{\alpha\in\mathrm{FL}}\tau_z^a$, for different thicknesses of the 
FePt layer. We present results for the situation where there is no SL (red squares), and for a Fe SL of respectively 2 (green 
squares) and 4 atomic planes (blue squares). For each SL thickness, the torkance shows little dependence on the thickness of
the FePt layer. When there is no SL this is attributed to the oscillatory behaviour without attenuation of the torkance profile 
as observed in Fig.~\ref{FePt:noSL}. In contrast, when a SL is present most of the torque resides at the first MgO/Fe interface 
so that the thickness of the FePt becomes irrelevant (see Fig.~\ref{FePt:FeSL}). Interestingly, when a SL is present the total 
torkance transferred into the Fe/MgO/Fe/FePt MTJ is larger than that of a simpler Fe/MgO/Fe MTJ with identical barrier (dashed
black line). This is no longer true when the SL is absent. Such finding means that the introduction of a Fe seed layer not only 
helps in achieving a better epitaxy during the growth but also facilitates a larger spin transfer torque. 

Figure~\ref{TotT+TMR}(b) shows the calculated TMR for each junction (in all cases the FePt layer comprises 4 layers) and a 
comparison with that of a Fe/MgO/Fe MTJ with an identical barrier. We observe that the junction with no Fe SL presents
the largest TMR, despite having the lowest torkance. This is unexpected, since in FePt bands with $\Delta_1$ symmetry, namely
those with the largest transmission across MgO, are present for both spin channels~\citesq{Taniguchi2008}. Such feature returns a 
predicted TMR for MTJs with FePt leads not exceeding 340\%~\citesq{Taniguchi2008}. However, here the situation is different since
in all our MTJs the leads are made of Fe, so that spin filtering is always in place. As such, in our case the addition of a FePt layer 
(or a complex Fe/FePt layer) changes the details of the spin-dependent scattering potential, but does not alter the main spin-filtering
mechanism at play in Fe/MgO/Fe junctions. Interestingly, as the thickness of the Fe SL gets larger, the value of the TMR is
reduced. 

\begin{figure*}[!t]
\begin{center}
	\includegraphics[width=17.78cm]{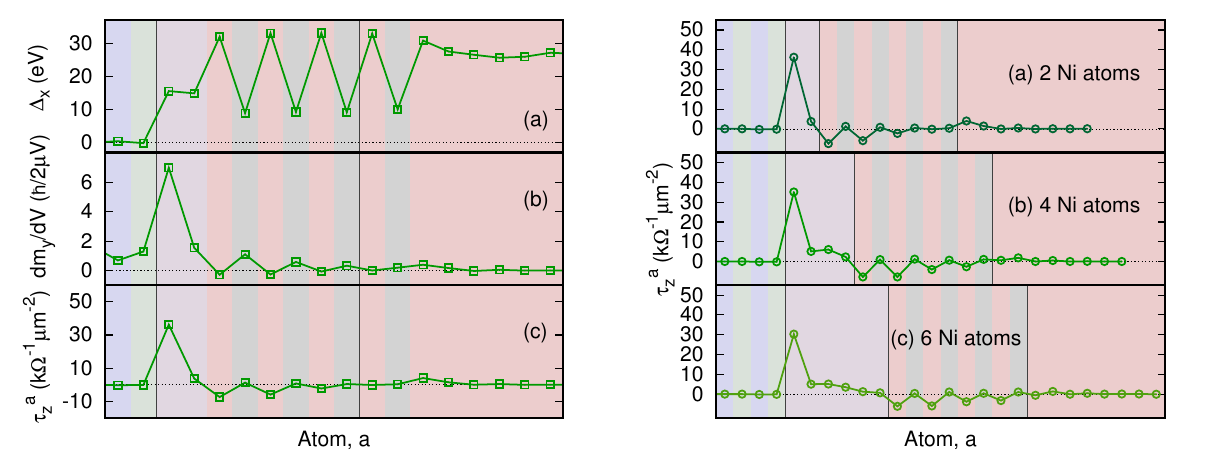}
\caption{(Color Online) Study of the torkance in a Fe/MgO/Ni/FePt/Fe junctions having a Ni seed layer of 2 (d), 4 (e) and 6 (f) 
monolayers. Left panel: the relevant components of (a) the exchange and correlation field, $\boldsymbol\Delta$, (b) the 
non-equilibrium spin density, $\mathrm{d}\boldsymbol{m}/\mathrm{d}V$ and (c) the torkance per unit of $\mu_\mathrm{B}/e$ 
and area, $\boldsymbol\tau$. Right Panel: comparison of the torkance per unit $\mu_\mathrm{B}/e$ and area of MTJs with 
with a seed layer comprising (d) 2, (e) 4 and (f) 6 Ni monolayers. The coloured background represents the different atomic
species: red for Fe, grey for Pt, blue for O, green for Mg, purple for Ni.}\label{Ni}
 \end{center}
\end{figure*}

So far the left electrode has been considered to be the fixed layer, namely the one producing the spin-polarised current. It is now 
interesting to look at the opposite case, namely the one where the electrons flux flows from the right-hand side to the left-hand side 
electrode. This is the situation where the FePt/Fe composite electrode acts as the fixed, current polarising, layer. Since in the right 
electrode the magnetisation is along the $z$ direction, the relevant torque in this case is $\tau_x$. This is presented in Fig.~\ref{RefL} 
for three representative junctions: (a) Fe/MgO/Fe, (b) Fe/MgO/FePt(4)/Fe, 
and (c) Fe/MgO/Fe(2)/FePt(4)/Fe, where the numbers in parentheses indicate the number of unit cells. Since in this geometry the 
current flows in the opposite direction than previously, we have plotted $-\tau_x$, namely the torque component that will lead to an 
alignment of the magnetisations of the fixed and free layers. The trend of $-\tau_x$ is in all cases analogous to that of $\tau_z$ for 
the Fe/MgO/Fe MTJ [see Fig.~\ref{FeMgO} (c)], namely the STT is peaked at the magnet/insulator interface and is negligible 
elsewhere. The only significant difference between the three MTJs is the reduction of approximately a factor two of the peak intensity 
for the Fe/MgO/FePt(4)/Fe stack [panel (b)]. 

\section{Discussion}

The results presented so far indicate that the STT (the torkance) varies strongly with the distance from the MgO interface, and 
that the details depend subtly on the specific layer structure. In general, Fe seems capable of absorbing a significant amount of angular 
momentum, so that only a few Fe monolayers are enough to make the STT decay sharply from the MgO interface. The main 
cause of such effect has to be found in the intense Fe exchange field. In fact, the strong exchange interaction in Fe relaxes the 
non-equilibrium spin density (the spin accumulation) toward the local direction of the magnetisation within a few atomic layers 
from the interface, so that there is little $\mathrm{d}\boldsymbol{m}/\mathrm{d}V$ away from the interface itself. In addition the 
exchange and correlation field remains almost constant within the Fe layer, resulting in a torque that persists little away from the 
interface with MgO.

In L$1_0$ FePt the alternating planes of Fe and Pt lead to a magnetisation texture that is non-uniform at the atomic scale.
In particular ${\bf \Delta}$ is small at the Pt sites so that the average exchange and correlation field is reduced with respect to 
that of the pure Fe case. As a consequence the spin accumulation can penetrate longer into the free layer so that the STT decays
less sharply. In order to further investigate the effects of the exchange field on the spatial decay of the torque we now consider 
a Ni seed layer since it has a much smaller moment, and thus exchange field, than Fe. The calculation has been simplified by 
maintaining the {\it bcc} structure and the lattice constant of Fe. As such our device stack does not correspond to a likely 
experimental situation but just serves the purpose of comparing the different seed layers. The atomic resolved torkance for a 
Fe/MgO/Ni/FePt/Fe stack with a Ni seed layer comprising 2, 4 and 6 atomic planes is shown in figure \ref{Ni}. 

As in the case of a Fe seed layer, the torque [panel (c)] is strongly peaked at the Ni/MgO interface, but now it does not decay 
entirely and thus a non-vanishing STT with an oscillatory behaviour persists into the FePt layer. A closer look at the profile
of ${\bf \Delta}$ across the junction [panel (a)] reveals that the exchange and correlation field in Ni is about half of that of
Fe [see Figure~\ref{FePt:FeSL}(a)]. As a consequence, in Ni the spin accumulation does not relax along the local direction of 
the magnetization as efficiently as in Fe, a fact that can be appreciated by comparing Fig.~\ref{Ni}(b) with Fig.~\ref{FePt:FeSL}(b).
Interestingly, the attenuation of the spin accumulation and thus of the torque is not complete even for relatively thick Ni
seed layers, as can be seen in panels (d) through (f). A second interesting observation concerns the phase of the
oscillations of the STT in the FePt layer. In fact for a junction where FePt is in direct contact with the MgO barrier, the torque
is positive at the Fe planes and negative (although rather small) at the Pt ones. The same behaviour, although with a much
reduced torque is observed for Fe intercalation (in the presence of a Fe seed layer). In contrast when the seed layer is made 
of Ni the sign of the STT on the FePt layer changes, becoming negative at the Fe planes and positive (although small) at the Pt 
ones. As a result the total integrated torque over the entire free layer (seed layer plus FePt) for Ni intercalation is two thirds than 
that obtained with Fe intercalation.

Finally, we wish to make a few general remarks on the spatial dependence of the STT. Macroscopic models combining the 
Landau-Lifshitz-Gilbert equation for the magnetisation dynamics with a diffusion model for the spin accumulation
\citesq{Abert2015,Wang2006} suggest that the spin accumulation is maximised in regions where there is a large magnetisation 
gradient, namely at interfaces. This is confirmed here at the microscopic level. In all cases investigated we find the maximum
spin accumulation, and hence torque, at the interface between the free layer and MgO regardless of the presence of a seed
layer. Furthermore, we also find an enhanced spin accumulation and torque at the second interface between the free layer
and the Fe lead, although this is small since the spin accumulation always decays in the free layer. The fine details of the
spin accumulation profile depend on how the entire stack responds to the application of an external bias. This in turn is
affected by the reorganisation in the occupation of the states around the Fermi surface, which is indeed a subtle effect.

In general a large exchange splitting causes the spin accumulation to relax faster along the local magnetisation direction. 
As such we expect the spin accumulation to decay more severely in the free layer of stacks where there is a large torque 
at the first few atomic layers in contact with the MgO barrier. This in turn depends on the strength of the exchange and 
correlation field, which in the LSDA can be written as
\begin{equation}
{\bf\Delta}^\mathrm{LSDA}({\bf r})=\frac{\delta E^\mathrm{LSDA}_\mathrm{XC}}{\delta\mu_\mathrm{B}{\bf m}({\bf r})}=
-\frac{\partial \epsilon_\mathrm{XC}}{\partial m({\bf r})}\:\frac{n({\bf r})}{\mu_\mathrm{B}}\:\frac{{\bf m}({\bf r})}{m({\bf r})}\:,
\end{equation}
where ${\bf m}({\bf r})$ is the local magnetisation vector, $m({\bf r})=|{\bf m}({\bf r})|$, $E^\mathrm{LSDA}_\mathrm{XC}$
is the LSDA exchange and correlation energy, $\epsilon_\mathrm{XC}$ is the exchange and correlation energy density of
the homogeneous electron gas, $n({\bf r})$ is the charge density and $\mu_\mathrm{B}$ the Bohr magneton. Crucially the 
LSDA ${\bf\Delta}$ is locally parallel to the magnetisation direction. As such one expects ${\bf\Delta}$ (and hence 
the torque) to change sign as the local magnetisation changes sign (as in the case of antiferromagnets). Furthermore 
one can show that $|{\bf\Delta}|\sim Im$, where $I$ is the Stoner parameter \citesq{Simoni2017}. This means that for similar
Stoner coupling the exchange and correlation field is more intense for materials presenting larger magnetization. This last 
feature explains the difference in ${\bf\Delta}$ and torque between the Fe and the Ni seed layer. In fact Fe and Ni have rather
similar Stoner parameter but their magnetization differ by more than a factor three.

\section{Conclusions}

In conclusion, we have calculated the STT acting upon the free ferromagnetic layer in a series of FePt-based magnetic tunnel junctions. 
For a simple Fe/MgO/Fe MTJ the torkance is peaked at the MgO interface and decays within 4 atomic planes. When the stack is modified
to include FePt [Fe/MgO/FePt/Fe] the torkance decays much slower and persists into the free layer up to at least 12 atomic planes. 
Such retention is associated to torkance oscillations at the length scale of the Fe-Pt plane separation. Since the lattice mismatch 
between MgO and FePt is large we have explored the option to intercalate a Fe seed layer at the interface between MgO and FePt. 
Also in this case the torkance is significant only at the first MgO/Fe interface and it vanishes in FePt. This is the result of the strong 
reduction of the spin accumulation beyond the Fe seed layer. Such strong attenuation appears to originate from the large exchange
and correlation field in Fe, which rapidly aligns the spin accumulation along the local direction of magnetization. Such hypothesis
is confirmed by calculations for the STT in some hypothetical MTJs incorporating a Ni seed layer. Since Ni has an exchange
and correlation field that is weaker than that of Fe, it is less effective at suppressing the spin accumulation (in absorbing angular 
momentum) and thus the attenuation of the torkance is weaker. All together our results suggest that the atomic and materials details 
of the MTJs stack play an important role in determining the total STT that a free layer can experience. This knowledge can help in 
designing stacks with maximal torkance, so that a reduction in the critical current for switching can be achieved. 

\section{Acknowledgements}

This work has been supported by the Science Foundation Ireland Principal Investigator award (grant no. 14/IA/2624 and 
16/US-C2C/3287) and TCHPC (Research IT, Trinity College Dublin). The authors wish to acknowledge the DJEI/DES/SFI/HEA 
Irish Centre for High-End Computing (ICHEC) for the provision of computational facilities and support.

\end{document}